# Tensor coupling and relativistic spin and pseudospin symmetries with the Hellmann potential

A. A. Rajabi, M. Hamzavi[*]

*Physics Department, Shahrood University of Technology, Shahrood, Iran*

[*]*Corresponding author: Tel.:+98 273 3395270, fax: +98 273 3395270*

[*]Email: majid.hamzavi@gmail.com

**Abstract**

The Hellmann potential is a superposition potential that consists of an attractive Coulomb potential $-a/r$ and a Yukawa potential $be^{-\delta r}/r$. By using the generalized parametric Nikiforov-Uvarov (NU) method, we have studied the approximate analytical solutions of the Dirac equation with the Hellmann potential including a Coulomb-like tensor potential for arbitrary spin-orbit quantum number $\kappa$ under the presence of exact spin and pseudo-spin (p-spin) symmetries. We show that tensor interaction removes degeneracies between spin and pseudospin doublets. As particular cases, we found the energy levels of non-relativistic case and also the pure Coulomb potential energy levels.

## 1- Introduction

The Dirac equation, which describes the motion of a spin-1/2 particle, has been used in solving many problems of nuclear and high-energy physics. The spin and the p-spin symmetries of the Dirac Hamiltonian had been discovered many years ago, however, these symmetries have recently been recognized empirically in nuclear and hadronic spectroscopes [1]. Within the framework of Dirac equation, p-spin symmetry used to feature the deformed nuclei and the super deformation to establish an effective shell-model [2-4], whereas spin symmetry is relevant for mesons [5]. The spin symmetry occurs when the scalar potential $S(r)$ is nearly equal to the vector potential $V(r)$ or equivalently $S(r) \approx V(r)$ and p-spin symmetry occurs when $S(r) \approx -V(r)$ [6,7].

The p-spin symmetry refers to a quasi-degeneracy of single nucleon doublets with non-relativistic quantum number $(n,l,j=l+1/2)$ and $(n-1,l+2,j=l+3/2)$, where $n$, $l$ and $j$ are single nucleon radial, orbital and total angular quantum numbers, respectively [8,9]. The total angular momentum is given by $j=\tilde{l}+\tilde{s}$, where $\tilde{l}=l+1$ pseudo-angular momentum and $\tilde{s}$ is p-spin angular momentum [10, 11]. Liang et al. [12] investigated the symmetries of the Dirac Hamiltonian and their breaking in realistic nuclei in the framework of perturbation theory. Guo [13] used the similarity renormalization group to transform the spherical Dirac operator into a diagonal form and then the upper (lower) diagonal element became an operator describing Dirac (anti-)particle, which holds the form of the Schrödinger-like operator with the singularity disappearing in every component. Chen and Guo [14] investigated the evolution toward the non-relativistic limit from the solutions of the Dirac equation by a continuous transformation of the Compton wavelength $\lambda$. Lu et al. [15] recently showed that the p-spin symmetry in single particle resonant states in nuclei is conserved when the attractive scalar and repulsive vector potentials have the same magnitude but opposite sign.

Tensor potentials were introduced into the Dirac equation with the substitution $\vec{p} \rightarrow \vec{p} - im\omega\beta.\hat{r}U(r)$ and a spin-orbit coupling is added to the Dirac Hamiltonian [16-17]. Lisboa et al. [18] have studied a generalized relativistic harmonic oscillator for spin-1/2 particles by considering a Dirac Hamiltonian that contains quadratic vector and scalar potentials together with a linear tensor potential, under the conditions of pseudospin and spin symmetry. Alberto et al. [19] studied the contribution of the isoscalar tensor coupling to the realization of pseudospin symmetry in nuclei. Akcay showed that the Dirac equation with scalar and vector quadratic potentials and a Coulomb-like tensor potential can be solved exactly [20], also, he exactly solved Dirac equation with tensor potential containing a linear and Coulomb-like terms, too [21]. Aydoğdu and Sever obtained exact solution of Dirac equation for the pseudoharmonic potential in the presence of linear tensor potential under the pseudospin symmetry and showed that tensor interactions remove all degeneracies between members of pseudospin doublets [22]. Ikhdair and Sever solved Dirac equation approximately for Hulthén potential including Coulomb-like tensor potential with arbitrary spin-orbit coupling number $\kappa$ under spin and pseudospin symmetry limit [10]. Aydoğdu and Sever solved approximately for the Woods-Saxon potential and a tensor potential with the arbitrary spin-orbit coupling quantum number $\kappa$ under pseudospin and spin symmetry [23]. Very recently, Hamzavi et al. gave exact solutions of the Dirac equation for Mie-type potential and position-dependent mass Coulomb potential with a Coulomb-like tensor potential [24,25] and pseudoharmonic potential with linear plus Coulomb-like tensor potential [26].

The superposition of the Coulomb plus Yukawa potential suggested by Hellmann is given by;

$$V(r) = -\frac{a}{r} + b\frac{e^{-\delta r}}{r}, \qquad (1)$$

where $a$ and $b$ are the strengths of the Coulomb and the Yukawa potentials, respectively, and $\delta$ is the screening parameter [27,28]. The Hellmann potential has been used by various authors to represent the electron-core [29,30] or the electron-ion [31,32] interaction. Varshni and Shukla [33] used this model potential for alkali hydride molecules. Das and Chakravarty [34] have proposed that such a potential is suitable for the study of inner-shell ionization problems. Adamowski [35] studied the bound-state energies of the Hellmann potential for various sets of values of $b$ and $\delta$ in a variational framework using ten variational parameters. Dutt et al [36] have also been investigated the bound-state energies as well as the wave functions of this potential using the large-N expansion technique. Hall and Katatbeh used potential envelopes method to analyze the bound state spectrum of the Schrödinger Hamiltonian with a Hellmann potential [37]. Ikhdair and Sever investigated energy levels of neutral atoms by applying an alternative perturbative scheme in solving the Schrödinger equation for the Yukawa potential model with a modified screening parameter [38]. Ikhdair and Sever also studied bound states of the Hellmann potential with arbitrary strength $b$ and screening parameter $\delta$ by using a perturbative approach [39]. Roy et al. studied the Hellman problem using a generalized pseudospectral method [40]. Nasser and Abdelmonem, using the J-matrix approach, studied the trajectories of the poles of the S-matrix for a Hellmann potential in the complex energy plane near the critical screening parameter [41].

The structure of the paper is as follows. In section 2, the parametric generalization of the NU method is displayed. In Section 3, in the context of spin and p-spin symmetry, we briefly introduce the Dirac equation with scalar and vector Hellmann potential for arbitrary spin-orbit quantum number $\kappa$. In the presence of the spin and p-spin symmetry, the approximate energy eigenvalue equations and corresponding two-component wave functions of the Dirac-Hellmann problem are obtained. The non-relativistic limit of the problem is discussed in this section too. Finally, our final concluding remarks are given in Section 4.

**2. Parametric NU Method**

This powerful mathematical tool solves second order differential equations. Let us consider the following differential equation [42-44]

$$\psi_n''(s) + \frac{\tilde{\tau}(s)}{\sigma(s)}\psi_n'(s) + \frac{\tilde{\sigma}(s)}{\sigma^2(s)}\psi_n(s) = 0, \tag{2}$$

where $\sigma(s)$ and $\tilde{\sigma}(s)$ are polynomials, at most of second degree, and $\tilde{\tau}(s)$ is a first-degree polynomial. To make the application of the NU method simpler and direct without need to check the validity of solution, we present a shortcut for the method. At first we write the general form of the Schrödinger-like equation (2) in a more general form as

$$\psi_n''(s) + \left(\frac{c_1 - c_2 s}{s(1-c_3 s)}\right)\psi_n'(s) + \left(\frac{-p_2 s^2 + p_1 s - p_0}{s^2(1-c_3 s)^2}\right)\psi_n(s) = 0, \tag{3}$$

satisfying the wave functions

$$\psi_n(s) = \phi(s) y_n(s). \tag{4}$$

Comparing (3) with its counterpart (2), we obtain the following identifications:

$$\tilde{\tau}(s) = c_1 - c_2 s, \quad \sigma(s) = s(1-c_3 s), \quad \tilde{\sigma}(s) = -p_2 s^2 + p_1 s - p_0, \tag{5}$$

Following the NU method [42], we obtain the bound-state energy condition [43,44]

$$c_2 n - (2n+1)c_5 + (2n+1)\left(\sqrt{c_9} + c_3\sqrt{c_8}\right) + n(n-1)c_3 + c_7 + 2c_3 c_8 + 2\sqrt{c_8 c_9} = 0, \tag{6}$$

resulting also in

$$\rho(s) = s^{c_{10}}(1-c_3 s)^{c_{11}}, \quad \phi(s) = s^{c_{12}}(1-c_3 s)^{c_{13}}, \quad c_{12} > 0, \ c_{13} > 0,$$

$$y_n(s) = P_n^{(c_{10}, c_{11})}(1-2c_3 s), \quad c_{10} > -1, \ c_{11} > -1, \tag{7a}$$

so that the wave function becomes

$$\psi_{nl}(s) = N_{nl} s^{c_{12}}(1-c_3 s)^{c_{13}} P_n^{(c_{10}, c_{11})}(1-2c_3 s). \tag{7b}$$

where $P_n^{(\mu,\nu)}(x)$, $\mu > -1$, $\nu > -1$, $x \in [-1,1]$ are Jacobi polynomials with parameters

$$c_4 = \frac{1}{2}(1-c_1), \qquad c_5 = \frac{1}{2}(c_2 - 2c_3),$$

$$c_6 = c_5^2 + p_2; \qquad c_7 = 2c_4c_5 - p_1,$$

$$c_8 = c_4^2 + p_0, \qquad c_9 = c_3(c_7 + c_3c_8) + c_6,$$

$$c_{10} = c_1 + 2c_4 + 2\sqrt{c_8} - 1 > -1, \qquad c_{11} = 1 - c_1 - 2c_4 + \frac{2}{c_3}\sqrt{c_9} > -1,\ c_3 \neq 0,$$

$$c_{12} = c_4 + \sqrt{c_8} > 0, \qquad c_{13} = -c_4 + \frac{1}{c_3}(\sqrt{c_9} - c_5) > 0,\ c_3 \neq 0, \qquad (8)$$

where $c_{12} > 0$, $c_{13} > 0$ and $s \in [0, 1/c_3]$, $c_3 \neq 0$.

In the rather more special case of $c_3 = 0$, the wave function (24) becomes

$$\lim_{c_3 \to 0} P_n^{(c_{10}, c_{11})}(1 - 2c_3 s) = L_n^{c_{10}}(c_{11} s), \quad \lim_{c_3 \to 0}(1 - c_3 s)^{c_{13}} = e^{c_{13} s},$$

$$\psi(s) = N s^{c_{12}} e^{c_{13} s} L_n^{c_{10}}(c_{11} s). \qquad (9)$$

where $L_n(x)$ is a Laguerre polynomial.

## 2. Dirac Equation including Tensor Coupling

The Dirac equation for fermionic massive spin-$\frac{1}{2}$ particles moving in an attractive scalar potential $S(r)$, a repulsive vector potential $V(r)$ and a tensor potential $U(r)$ is $[\hbar = c = 1]$

$$[\vec{\alpha}.\vec{p} + \beta(M + S(r)) - i\beta\vec{\alpha}.\hat{r}U(r)]\psi(\vec{r}) = [E - V(r)]\psi(\vec{r}), \qquad (10)$$

where $E$ is the relativistic energy of the system, $\vec{p} = -i\vec{\nabla}$ is the three-dimensional momentum operator and $M$ is the mass of the fermionic particle [45]. $\vec{\alpha}$ and $\beta$ are the $4 \times 4$ usual Dirac matrices give as

$$\vec{\alpha} = \begin{pmatrix} 0 & \vec{\sigma} \\ \vec{\sigma} & 0 \end{pmatrix}, \quad \beta = \begin{pmatrix} I & 0 \\ 0 & -I \end{pmatrix}, \qquad (11)$$

where $I$ is $2 \times 2$ unitary matrix and $\vec{\sigma}$ are three-vector spin matrices

$$\sigma_1 = \begin{pmatrix} 0 & 1 \\ 1 & 0 \end{pmatrix}, \quad \sigma_2 = \begin{pmatrix} 0 & -i \\ i & 0 \end{pmatrix}, \quad \sigma_3 = \begin{pmatrix} 1 & 0 \\ 0 & -1 \end{pmatrix}. \qquad (12)$$

The total angular momentum operator $\vec{J}$ and spin-orbit $K = (\vec{\sigma}.\vec{L} + 1)$, where $\vec{L}$ is orbital angular momentum, of the spherical nucleons commute with the Dirac Hamiltonian. The eigenvalues of spin-orbit coupling operator are $\kappa = \left(j + \frac{1}{2}\right) > 0$ and $\kappa = -\left(j + \frac{1}{2}\right) \langle 0$ for unaligned spin $j = l - \frac{1}{2}$ and the aligned spin $j = l + \frac{1}{2}$, respectively. $(H^2, K, J^2, J_z)$ can be taken as the complete set of the conservative quantities. Thus, the spinor wave functions can be classified according to their angular momentum $j$, spin-orbit quantum number $\kappa$, and the radial quantum number $n$, and can be written as follows

$$\psi_{n\kappa}(\vec{r}) = \begin{pmatrix} f_{n\kappa}(\vec{r}) \\ g_{n\kappa}(\vec{r}) \end{pmatrix} = \begin{pmatrix} \dfrac{F_{n\kappa}(r)}{r} Y^l_{jm}(\theta, \varphi) \\ i \dfrac{G_{n\kappa}(r)}{r} Y^{\tilde{l}}_{jm}(\theta, \varphi) \end{pmatrix}, \tag{13}$$

where $f_{n\kappa}(\vec{r})$ is the upper (large) component and $g_{n\kappa}(\vec{r})$ is the lower (small) component of the Dirac spinors. $Y^l_{jm}(\theta, \varphi)$ and $Y^{\tilde{l}}_{jm}(\theta, \varphi)$ are spin and pseudospin spherical harmonics, respectively, and $m$ is the projection of the angular momentum on the $z$-axis. Substituting Eq. (13) into Eq. (10) and using the following relations

$$(\vec{\sigma}.\vec{A})(\vec{\sigma}.\vec{B}) = \vec{A}.\vec{B} + i\vec{\sigma}.(\vec{A} \times \vec{B}), \tag{14a}$$

$$(\vec{\sigma}.\vec{P}) = \vec{\sigma}.\hat{r}\left(\hat{r}.\vec{P} + i\frac{\vec{\sigma}.\vec{L}}{r}\right). \tag{14b}$$

With the following properties

$$\begin{aligned}
(\vec{\sigma}.\vec{L}) Y^{\tilde{l}}_{jm}(\theta, \phi) &= (\kappa - 1) Y^{\tilde{l}}_{jm}(\theta, \phi), \\
(\vec{\sigma}.\vec{L}) Y^l_{jm}(\theta, \phi) &= -(\kappa - 1) Y^l_{jm}(\theta, \phi), \\
(\vec{\sigma}.\hat{r}) Y^{\tilde{l}}_{jm}(\theta, \phi) &= -Y^l_{jm}(\theta, \phi), \\
(\vec{\sigma}.\hat{r}) Y^l_{jm}(\theta, \phi) &= -Y^{\tilde{l}}_{jm}(\theta, \phi),
\end{aligned} \tag{15}$$

one obtains two coupled differential equations for upper and lower radial wave functions $F_{n\kappa}(r)$ and $G_{n\kappa}(r)$ as

$$\left(\frac{d}{dr} + \frac{\kappa}{r} - U(r)\right) F_{n\kappa}(r) = (M + E_{n\kappa} - \Delta(r)) G_{n\kappa}(r), \tag{16a}$$

$$\left(\frac{d}{dr} - \frac{\kappa}{r} + U(r)\right) G_{n\kappa}(r) = (M - E_{n\kappa} + \Sigma(r)) F_{n\kappa}(r). \tag{16b}$$

where

$$\Delta(r) = V(r) - S(r), \tag{17a}$$

$$\Sigma(r) = V(r) + S(r). \tag{17b}$$

Eliminating $F_{n\kappa}(r)$ and $G_{n\kappa}(r)$ from Eqs. (17), we obtain the following two Schrödinger-like differential equations for the upper and lower radial spinor components, respectively:

$$\left[\frac{d^2}{dr^2} - \frac{\kappa(\kappa+1)}{r^2} + \frac{2\kappa}{r}U(r) - \frac{dU(r)}{dr} - U^2(r)\right]F_{n\kappa}(r)$$
$$+ \frac{\frac{d\Delta(r)}{dr}}{M + E_{n\kappa} - \Delta(r)}\left(\frac{d}{dr} + \frac{\kappa}{r} - U(r)\right)F_{n\kappa}(r) \tag{18}$$
$$= \left[(M + E_{n\kappa} - \Delta(r))(M - E_{n\kappa} + \Sigma(r))\right]F_{n\kappa}(r),$$

$$\left[\frac{d^2}{dr^2} - \frac{\kappa(\kappa-1)}{r^2} + \frac{2\kappa}{r}U(r) + \frac{dU(r)}{dr} - U^2(r)\right]G_{n\kappa}(r)$$
$$+ \frac{\frac{d\Sigma(r)}{dr}}{M - E_{n\kappa} + \Sigma(r)}\left(\frac{d}{dr} - \frac{\kappa}{r} + U(r)\right)G_{n\kappa}(r) \tag{19}$$
$$= \left[(M + E_{n\kappa} - \Delta(r))(M - E_{n\kappa} + \Sigma(r))\right]G_{n\kappa}(r)$$

where $\kappa(\kappa-1) = \tilde{l}(\tilde{l}+1)$ and $\kappa(\kappa+1) = l(l+1)$. The quantum number $\kappa$ is related to the quantum numbers for spin symmetry $l$ and pseudospin symmetry $\tilde{l}$ as

$$\kappa = \begin{cases} -(l+1) = -(j+\frac{1}{2}) & (s_{1/2}, p_{3/2}, etc.) \quad j = l+\frac{1}{2}, \quad \text{aligned spin}\,(\kappa\langle 0) \\ +l = +(j+\frac{1}{2}) & (p_{1/2}, d_{3/2}, etc.) \quad j = l-\frac{1}{2}, \quad \text{unaligned spin}\,(\kappa\rangle 0), \end{cases}$$

and the quasidegenerate doublet structure can be expressed in terms of a pseudospin angular momentum $\tilde{s} = 1/2$ and pseudo-orbital angular momentum $\tilde{l}$, which is defined as

$$\kappa = \begin{cases} -\tilde{l} = -(j+\frac{1}{2}) & (s_{1/2}, p_{3/2}, etc.) \quad j = \tilde{l}-\frac{1}{2}, \quad \text{alinged pseudospin}\,(\kappa\langle 0) \\ +(\tilde{l}+1) = +(j+\frac{1}{2}) & (d_{3/2}, f_{5/2}, etc.) \quad j = \tilde{l}+\frac{1}{2}, \quad \text{unaligned spin}\,(\kappa\rangle 0), \end{cases}$$

where $\kappa = \pm 1, \pm 2, \ldots$. For example, $(1s_{1/2}, 0d_{3/2})$ and $(1p_{3/2}, 0f_{5/2})$ can be considered as pseudospin doublets.

## 2.1. Spin Symmetric Limit

In the spin symmetric limitation, $\frac{d\Delta(r)}{dr} = 0$ or $\Delta(r) = C_s = $ constant [7, 46-48], then Eq. (16) with $\Sigma(r)$ as Hellmann potential becomes

$$\left[\frac{d^2}{dr^2} - \frac{\eta_\kappa(\eta_\kappa - 1)}{r^2} - \gamma\left(-\frac{a}{r} + b\frac{e^{-\delta r}}{r}\right) - \beta^2\right] F_{n\kappa}(r) = 0, \tag{20a}$$

$$\eta_\kappa = \kappa + H + 1, \; \gamma = M + E_{n\kappa} - C_s \text{ and } \beta^2 = (M - E_{n\kappa})(M + E_{n\kappa} - C_s). \tag{20b}$$

where $\kappa = l$ and $\kappa = -l - 1$ for $\kappa < 0$ and $\kappa > 0$, respectively. The Schrödinger-like equation (20a) that results from the Dirac equation is a second order differential equation containing a spin-orbit centrifugal term $\eta_\kappa(\eta_\kappa + 1)r^{-2}$ which has a strong singularity at $r = 0$, and needs to be treated very carefully while performing the approximation. In absence of tensor interaction, equation (20a) has an exact rigorous solution only for the states with $\kappa = -1$ because of the existence of the centrifugal term $\kappa(\kappa + 1)/r^2$. However, when this term is taken into account, the corresponding radial Dirac equation can no longer be solved in a closed form and it is necessary to resort to approximate methods. Over the last few decades several schemes have been used to calculate the energy spectrum. The main idea of these schemes relies on using different approximations of the spin-orbit centrifugal coupling term $\eta_\kappa(\eta_\kappa + 1)/r^2$. So we need to perform a new approximation for the spin-orbit term as a function of the tPT potential parameters. Therefore, we resort to use an appropriate approximation scheme to deal with the centrifugal potential term as

$$\frac{1}{r^2} \approx \frac{\delta^2}{(1 - e^{-\delta r})^2}, \tag{21}$$

or equivalently

$$\frac{1}{r} \approx \frac{\delta}{1 - e^{-\delta r}}, \tag{22}$$

which is valid for $\delta r \ll 1$ [49]. Therefore, the Hellmann potential in (1) reduces to [50-52]

$$V(r) \approx -\frac{\delta a}{1 - e^{-\delta r}} + \frac{\delta b e^{-\delta r}}{1 - e^{-\delta r}}. \tag{23}$$

To see the accuracy of our approximation, we plotted the Hellmann potential (1) and its approximation (23) with parameters $a=2$, $b=-4$ and $\delta=0.01$ [39], in Figure 1. Thus, employing such an approximation scheme, we can then write Eq. (20a) as:

$$\left[\frac{d^2}{dr^2}-\eta_\kappa(\eta_\kappa-1)\frac{\delta^2}{(1-e^{-\delta r})^2}-\gamma\left(-\frac{\delta a}{1-e^{-\delta r}}+\frac{\delta b e^{-\delta r}}{1-e^{-\delta r}}\right)-\beta^2\right]F_{n\kappa}(r)=0. \qquad (24)$$

Followed by making a new change of variables $s=e^{-\delta r}$, this allows us to decompose the spin-symmetric Dirac equation (20) into the Schrödinger-type equation satisfying the upper-spinor component $F_{n,\kappa}(s)$,

$$\left\{\frac{d^2}{ds^2}+\frac{1-s}{s(1-s)}\frac{d}{ds}+\frac{1}{s^2(1-s)^2}\left[-As^2+Bs-C\right]\right\}F_{n,\kappa}(s)=0,$$

$$A=-\frac{\gamma b}{\delta}+\frac{\beta^2}{\delta^2},$$

$$B=-\frac{\gamma a}{\delta}-\frac{\gamma b}{\delta}+\frac{2\beta^2}{\delta^2},$$

$$C=\eta_\kappa(\eta_\kappa-1)-\frac{\gamma a}{\delta}+\frac{\beta^2}{\delta^2}. \qquad (25)$$

where $F_{n\kappa}(r)\equiv F_{n,\kappa}(s)$ has been used. If the above equation is compared with Eq. (3), we can obtain the specific values for constants $c_i$ ($i=1,2,3$) as

$$c_1=1, \ c_2=1 \text{ and } c_3=1. \qquad (26)$$

In order to obtain the bound state solutions of Eq. (25), it is necessary to calculate the remaining parametric constants, that is, $c_i$ ($i=4,5,...,13$) by means of the relation (8). Their specific values are displayed in table 1 for the relativistic Hellmann potential model. Further, using these constants along with (6), we can readily obtain the energy eigenvalue equation for the Dirac-Hellmann problem as

$$\eta_\kappa(\eta_\kappa+1)+(n+\eta_\kappa+1)^2+(2n+\eta_\kappa+1/2)\sqrt{\eta_\kappa(\eta_\kappa-1)-\frac{\gamma a}{\delta}+\frac{\beta^2}{\delta^2}}-\frac{\gamma}{\delta}(b-a)=0 \qquad (27)$$

To show the procedure of determining the energy eigenvalues from Eq. (27), we take a set of physical parameter values, $M=5\,\text{fm}^{-1}$, $a=1$, $b=-4$, $C_s=5.5\,\text{fm}^{-1}$ and $\delta=0.01$ [28]. In table 2, we present the energy spectrum for the spin symmetric case. Obviously, the pairs $(np_{1/2},np_{3/2})$, $(nd_{3/2},nd_{5/2})$, $(nf_{5/2},nf_{7/2})$, $(ng_{7/2},ng_{9/2})$, and so on are degenerate states. Also, we can see that tensor interaction remove degeneracies between spin doublets.

In Figure 2, we have investigated the effect of the tensor potential on the spin doublet splitting by considering the following pairs of orbital: $(1p_{3/2}, 1p_{1/2})$, $(1f_{7/2}, 1f_{5/2})$. We observe that in the case of $T = 0$ (no tensor interaction), members of spin doublets have same energy. However, in the presence of the tensor potential $T \neq 0$, these degeneracies are removed. We can also see that spin doublet splitting increases with increasing $T$. The reason is that term $2\kappa T$ gives different contributions to each level in the spin doublet because $T$ takes different values for each state in the spin doublet. On the other hand, in order to establish the upper-spinor component of the wave functions $F_{n,\kappa}(r)$, namely, Eq. (20a), the relations (7a) and (7b) are used. Firstly, we find the first part of the wave function as

$$\rho(s) = s^{2\sqrt{C}} (1-s)^{2\eta_\kappa - 1}, \tag{28}$$

Secondly, we calculate the weight function as

$$\phi(s) = s^{\sqrt{C}} (1-s)^{\eta_\kappa} \tag{29}$$

which gives the second part of the wave function as

$$y_n(s) = P_n^{(2\sqrt{C}, 2\kappa+1)}(1-2s) \tag{30}$$

where $P_n^{(a,b)}(y)$ are the orthogonal Jacobi polynomials. Finally the upper spinor component for arbitrary $\kappa$ can be found through the relation (7b)

$$F_{n\kappa}(s) = N_{n\kappa} s^{\sqrt{C}} (1-s)^{\eta_\kappa} P_n^{(2\sqrt{C}, 2\eta_\kappa - 1)}(1-2s), \tag{31}$$

or

$$F_{n\kappa}(r) = N_{n\kappa} e^{-\delta\sqrt{C} r} (1 - e^{-\delta r})^{\eta_\kappa} P_n^{(2\sqrt{C}, 2\eta_\kappa - 1)}(1 - 2e^{-\delta r}), \tag{32}$$

where $N_{n\kappa}$ is the normalization constant. Further, the lower-spinor component of the wave function can be calculated by using

$$G_{n\kappa}(r) = \frac{1}{M + E_{n\kappa} - C_s} \left( \frac{d}{dr} + \frac{\kappa}{r} - U(r) \right) F_{n\kappa}(r), \tag{33}$$

where $E \neq -M + C_s$ and only positive energy states do exist.

## 2.2. P-spin Symmetric Limit

Ginocchio showed that there is p-spin symmetry in case when the relationship between the vector potential and the scalar potential is given by $V(r) = -S(r)$ [7,46-48]. Further, Meng et al. showed that if $\frac{d[V(r)+S(r)]}{dr} = \frac{d\Sigma(r)}{dr} = 0$, then $\Sigma(r) = C_{ps} =$ constant, for which the p-spin symmetry is exact in the Dirac equation. Thus, choosing the $\Delta(r)$ as Hellmann potential and $U(r)$ as Coulomb potential, Eq. (19) under this symmetry becomes

$$\left[\frac{d^2}{dr^2} - \frac{\lambda_\kappa(\lambda_\kappa - 1)}{r^2} - \tilde{\gamma}\left(-\frac{a}{r} + b\frac{e^{-\delta r}}{r}\right) - \tilde{\beta}^2\right]G_{n\kappa}(r) = 0, \tag{34a}$$

$$\lambda_\kappa = \kappa + H \quad \tilde{\gamma} = E_{n\kappa} - M - C_{ps} \quad \text{and} \quad \tilde{\beta}^2 = (M + E_{n\kappa})(M - E_{n\kappa} + C_{ps}), \tag{34b}$$

where $\kappa = -\tilde{l}$ and $\kappa = \tilde{l} + 1$ for $\kappa < 0$ and $\kappa > 0$, respectively. Employing the new approximation, the p-spin Dirac equation (35a) can be written as

$$\left[\frac{d^2}{dr^2} - \lambda_\kappa(\lambda_\kappa - 1)\frac{\delta^2}{(1-e^{-\delta r})^2} - \tilde{\gamma}\left(-\frac{\delta a}{1-e^{-\delta r}} + \frac{\delta b e^{-\delta r}}{1-e^{-\delta r}}\right) - \tilde{\beta}^2\right]G_{n\kappa}(r) = 0. \tag{35}$$

To avoid repetition, the negative energy solution of Eq. (35), the p-spin symmetric case can be readily obtained directly via the spin symmetric solution throughout the following parametric mappings:

$F_{n\kappa}(r) \leftrightarrow G_{n\kappa}(r), \quad \kappa \to \kappa - 1, \quad V(r) \to -V(r) \text{ ( i.e., } V_1 \to -V_1, \; V_2 \to -V_2 \text{)},$

$E_{n\kappa} \to -E_{n\kappa}, \quad C_s \to -C_{ps}. \tag{36}$

Following the previous procedure, one can obtain the p-spin symmetric energy equation as

$$\lambda_\kappa(\lambda_\kappa - 1) + (n + \lambda_\kappa - 1)^2 + (2n + \lambda_\kappa + 1/2)\sqrt{\lambda_\kappa(\lambda_\kappa - 1) - \frac{\tilde{\gamma}a}{\delta} + \frac{\tilde{\beta}^2}{\delta^2}} - \frac{\tilde{\gamma}}{\delta}(b-a) = 0 \tag{37}$$

And also, by using transformation (36), the corresponding lower wave function can easily be obtained from Eq. (32).

In table 3, we give the numerical results for the p-spin symmetric case. In this case, we take the set of parameter values, $M = 5\,\text{fm}^{-1}$, $a = -1$, $b = 4$, $c_{ps} = -5.5\,\text{fm}^{-1}$ and $\delta = 0.01$ [38]. We observe the degeneracy in the following doublets $(1s_{1/2}, 0d_{3/2})$, $(1p_{3/2}, 0f_{5/2})$, $(1d_{5/2}, 0g_{7/2})$, $(1f_{7/2}, 0h_{9/2})$, and so on. Thus, each pair is considered as p-spin doublet and has negative energy. In table 3, we can see that tensor interaction remove degeneracies between spin doublets. Also, in the presence of the p-spin symmetry, only negative energy states do exist.

In Figure 3, we have investigated the effect of the tensor potential on the p-spin doublet splitting by considering the following pairs of orbital: $(1d_{5/2}, 0g_{7/2})$, $(2f_{7/2}, 1h_{9/2})$ and one can observe that the results obtained in the p-spin symmetric limit resemble the ones observed in the spin symmetric limit.

### 2.3. The non-relativistic limiting case

In this section, we study the energy eigenvalue equation (27) and upper-spinor component of wave function (32) of the Dirac-Hellmann problem under the non-relativistic limits: $H = 0$, $C_s = 0$, $\kappa \to l$, $E_{n\kappa} - M \simeq E_{nl}$ and $M + E_{n\kappa} \simeq 2m$. Thereby, we obtain the energy equation of the Schrödinger equation with any arbitrary orbital state for the Hellmann potential as

$$E_{nl} = -\frac{\delta^2}{2m}\left[\left(\frac{\frac{2m}{\delta}(a-b)-(n+l+1)^2-l(l+1)}{2(n+l+1)}\right)^2 - l(l+1) + \frac{2ma}{\delta}\right] \qquad (38)$$

In table 4, we obtained the energy eigenvalues of the Hellmann potential for various states by the presented method and compared with other results obtained by a perturbative treatment method [39]. In table 4, we reported the energy eigenvalues (in units $\hbar = 2m = 1$) of the Hellmann potential in for various values of the screening parameter while the strength of the Yukawa potential is $b = \pm 1, -2$ and $-4$. Also, the radial functions can be obtained as

$$R_{nl}(r) = N_{nl} e^{-\sqrt{-2m(E+\delta a)+l(l+1)\delta^2}\, r} \left(1-e^{-\delta r}\right)^{l+1} P_n^{(2\sqrt{\frac{-2m}{\delta^2}(E+\delta a)+l(l+1)}, 2l+1)}\left(1-2e^{-\delta r}\right). \qquad (39)$$

where $N_{nl}$ is normalization constant. In Figure 4, we plotted the normalized radial wave functions of the Hellmann potential for various states.

Finally, when the screening parameter $\delta$ approaches zero and also $b = 0$, the potential (1) reduces to a Coulomb potential. Thus, in this limit the energy eigenvalues of (38) become the energy levels of the pure Coulomb interaction, i.e.

$$E_{n,l\,Coulomb} = -\frac{1}{2}m\frac{a^2}{n'^2} \qquad (40)$$

where $n' = n + l + 1$ [52].

## 4. Concluding Remarks

In this work, we have studied the bound state solutions of the Dirac equation with Hellmann and Coulomb-like tensor potentials for any spin-orbit quantum number $\kappa$. By making an appropriate approximation to deal with the spin-orbit centrifugal (pseudo-centrifugal) coupling term, we have obtained the approximate energy eigenvalue equation and the unnormalized two components of the radial wave functions expressed in terms of the Jacobi polynomials using the NU method. It is found that tensor interaction removes degeneracies between each pairs of pseudospin or spin doublets. We obtained the non-relativistic case of the problem and also found the energy levels of the familiar pure Coulomb potential when the screening parameter goes to zero.

**References**

[1] J. N. Ginocchio, Phys. Rep. **414** (4-5) (2005) 165.

[2] A. Bohr, I. Hamamoto and B. R. Mottelson, Phys. Scr. **26** (1982) 267.

[3] J. Dudek, W. Nazarewicz, Z. Szymanski and G. A. Leander, Phys. Rev. Lett. **59** (1987) 1405.

[4] D. Troltenier, C. Bahri and J. P. Draayer, Nucl. Phys. A **586** (1995) 53.

[5] P. R. Page, T. Goldman and J. N. Ginocchio, Phys. Rev. Lett. **86** (2001) 204.

[6] J. N. Ginocchio, A. Leviatan, J. Meng, and S. G. Zhou, Phys. Rev. C **69** (2004) 034303.

[7] J. N. Ginocchio, Phys. Rev. Lett. **78** (3) (1997) 436.

[8] K. T. Hecht and A. Adler, Nucl. Phys. A **137** (1969) 129.

[9] A. Arima, M. Harvey and K. Shimizu, Phys. Lett. B **30** (1969) 517.

[10] S. G. Zhou, J. Meng, and P. Ring, Phys. Rev. Lett. 91 (2003) 262501.

[11] X. T. He, S. G. Zhou, J. Meng, E. G. Zhao, and W. Scheid, Euro. Phys. J. A 28 (2006) 265.

[12] H. Liang, P. Zhao, Y. Zhang, J. Meng and N. V. Giai, Phys Rev C **83** (2011) 041301(R)

[13] J.-Y. Guo, Phys Rev C **85** (2012) 021302(R)

[14] S.-W. Chen and J.-Y. Guo, Phys Rev C **85** (2012) 054312

[15] B.-N. Lu, E.-G. Zhao and S.-G. Zhou, Phys. Rev. Lett. **109** (2012) 072501

[16] M. Moshinsky and A. Szczepanika, J. Phys. A: Math. Gen. **22** (1989) L817.

[17] V. I. Kukulin, G. Loyla and M. Moshinsky, Phys. Lett. A **158** (1991) 19.

[18] R. Lisboa, M. Malheiro, A. S. de Castro, P. Alberto, M. Fiolhais, Phys. Rev. C **69** (2004) 024319.

[19] P. Alberto, R. Lisboa, M. Malheiro and A. S. de Castro, Phys. Rev. C **71** (2005) 034313.

[20] H. Akcay, Phys. Lett. A **373** (2009) 616.

[21] H. Akcay, J. Phys. A: Math. Theor. **40** (2007) 6427.

[22] O. Aydoğdu and R. Sever, Few-Body Syst. **47** (2010) 193.

[23] O. Aydoğdu and R. Sever, Eur. Phys. J. A **43** (2010) 73.

[24] M. Hamzavi, A. A. Rajabi, H. Hassanabadi, Few-Body Syst. **48** (2010) 171.

[25] M. Hamzavi, A. A. Rajabi and H. Hassanabadi, Phys. Lett. A **374** (2010) 4303.

[26] M. Hamzavi, A. A. Rajabi, H. Hassanabadi, Int. J. Mod. Phys. A **26** (2011) 1363.

[27] H. Hellmann, Acta Physicochim. URSS 1 (1935) 913; Acta Physicochim. URSS 4 (1936) 225; Acta Physicochim. URSS 4 (1936) 324; J. Chem. Phys. 3 (1935) 61.

[28] H. Hellmann, W. Kassatotchkin, Acta Physicochim. URSS 5 (1936) 23; H. Hellmann, W. Kassatotchkin, J. Chem. Phys. 4 (1936) 324.

[29] P. Gombas, Die Statistische Theorie des Atoms und ihre Anwendungen, Springer, Berlin, 1949, p. 304.

[30] J. Callaway, Phys. Rev. 112 (1958) 322; G. J. Iafrate, J. Chem. Phys. 45 (1966) 1072; J. Callaway, P. S. Laghos, Phys. Rev. 187 (1969) 192; J. McGinn, J. Chem. Phys. 53 (1970) 3635.

[31] V. K. Gryaznov, Zh. Eksp Teor. Fiz. 78 (1980) 573, [Sov. Phys.-JETP 51 (1980) 288].

[32] V. A. Alekseev, V .E. Fortov, I. T. Yakubov, Usp. Fiz. Nauk 139 (1983) 193, [Sov. Phys.-Usp. 26 (1983) 99].

[33] Y. P. Varshni, R.C. Shukla, Rev. Mod. Phys. 35 (1963) 130.

[34] J. N. Das, S. Chakravarty, Phys. Rev. A 32 (1985) 176.

[35] J. Adamowski, Phys. Rev. A 31 (1985) 43.

[36] R. Dutt, U. Mukherji, Y. P. Varshni, Phys. Rev. A 34 (1986) 777.

[37] R. L. Hall, Q. D. Katatbeh, Physics Letters A 287 (2001) 183.

[38] S. M. Ikhdair, R. Sever, Int J. Mod. Phys. A 21 (2006) 6465.

[39] S. M. Ikhdair, R. Sever, J. Mol. Struct.: THEOCHEM 809 (2007) 103.

[40] A. K. Roy, A. F. Jalbout, E. I. Proynov, J. Math. Chem. 44 (2008) 260.

[41] I. Nasser1, and M. S. Abdelmonem, Phys. Scr. 83 (2011) 055004.

[42] A. F. Nikiforov and V. B. Uvarov: Special Functions of Mathematical Physic, Birkhausr, Berlin, 1988.

[43] C. Tezcan and R. Sever, Int. J. Theor. Phys. 48 (2009) 337.

[44] S. M. Ikhdair, Int. J. Mod. Phys. C 20 (1) (2009) 25.

[45] W. Greiner, Relativistic Quantum Mechanics, Wave Equations, Third Edition, Springer, 2000.

[46] J. N. Ginocchio, Phys. Rep. **315** (1999) 231.

[47] J. Meng, K. Sugawara-Tanabe, S. Yamaji and A. Arima, Phys. Rev. C **59** (1999) 154.

[48] J. Meng, K. Sugawara-Tanabe, S. Yamaji, P. Ring and A. Arima, Phys. Rev. C **58** (1998) R628.

[49]. R. L. Greene and C. Aldrich, Phys. Rev. A 14 (1976) 2363.

[50] M. R. Setare and S. Haidari, Phys. Scr. 81 (2010) 065201.

[51] O. Aydoğdu and R. Sever, Phys. Scr. 84 (2011) 025005.

[52] M. Hamzavi, M. Movahedi, K.-E. Thylwe and A. A. Rajabi, Chin. Phys. Lett. 29 (2012) 080302.

**Table 1.** The specific values for the parametric constants necessary for the energy eigenvalues and eigenfunctions

| constant | Analytic value |
|----------|----------------|
| $c_4$ | $0$ |
| $c_5$ | $-\dfrac{1}{2}$ |
| $c_6$ | $\dfrac{1}{4}+A$ |
| $c_7$ | $-B$ |
| $c_8$ | $C$ |
| $c_9$ | $\left(\eta_\kappa-\dfrac{1}{2}\right)^2$ |
| $c_{10}$ | $2\sqrt{C}$ |
| $c_{11}$ | $2\eta_\kappa-1$ |
| $c_{12}$ | $\sqrt{C}$ |
| $c_{13}$ | $\eta_\kappa$ |

Table 2. The bound state energy eigenvalues in unit of $fm^{-1}$ of the spin symmetry Hellmann potential for several values of $n$ and $\kappa$.

| $l$ | $n, \kappa < 0$ | $(l, j = l+1/2)$ | $E_{n,\kappa<0}$ | $E_{n,\kappa<0}$ | $n, \kappa > 0$ | $(l, j = l-1/2)$ | $E_{n,\kappa>0}$ | $E_{n,\kappa>0}$ |
|---|---|---|---|---|---|---|---|---|
| 1 | 0, -2 | $0p_{3/2}$ | 1.122753084 | 2.266823746 | 0, 1 | $0p_{1/2}$ | 3.174420713 | 2.266823746 |
| 2 | 0, -3 | $0d_{5/2}$ | 2.266823746 | 3.174420713 | 0, 2 | $0d_{3/2}$ | 3.760219205 | 3.174420713 |
| 3 | 0, -4 | $0f_{7/2}$ | 3.174420713 | 3.760219205 | 0, 3 | $0f_{5/2}$ | 4.127994487 | 3.760219205 |
| 4 | 0, -5 | $0g_{9/2}$ | 3.760219205 | 4.127994487 | 0, 4 | $0g_{7/2}$ | 4.364846559 | 4.127994487 |
| 1 | 1, -2 | $1p_{3/2}$ | 2.261929071 | 3.167838743 | 1, 1 | $1p_{1/2}$ | 3.753448611 | 3.167838743 |
| 2 | 1, -3 | $1d_{5/2}$ | 3.167838743 | 3.753448611 | 1, 2 | $1d_{3/2}$ | 4.121562668 | 3.753448611 |
| 3 | 1, -4 | $1f_{7/2}$ | 3.753448611 | 4.121562668 | 1, 3 | $1f_{5/2}$ | 4.358895657 | 4.121562668 |
| 4 | 1, -5 | $1g_{9/2}$ | 4.121562668 | 4.358895657 | 1, 4 | $1g_{7/2}$ | 4.517787398 | 4.358895657 |

**Table 3.** The bound state energy eigenvalues in unit of $fm^{-1}$ of the p-spin symmetry Hellmann potential for several values of $n$ and $\kappa$.

| $\tilde{l}$ | $n, \kappa < 0$ | $(l, j)$ | $E_{n,\kappa<0}$ | $E_{n,\kappa<0}$ | $n-1, \kappa > 0$ | $(l+2, j+1)$ | $E_{n-1,\kappa>0}$ | $E_{n-1,\kappa>0}$ |
|---|---|---|---|---|---|---|---|---|
| 1 | 1, -1 | $1s_{1/2}$ | −2.261929071 | −3.167838743 | 0, 2 | $0d_{3/2}$ | −3.753448611 | −3.167838743 |
| 2 | 1, -2 | $1p_{3/2}$ | −3.167838743 | −3.753448611 | 0, 3 | $0f_{5/2}$ | −4.121562668 | −3.753448611 |
| 3 | 1, -3 | $1d_{5/2}$ | −3.753448611 | −4.121562668 | 0, 4 | $0g_{7/2}$ | −4.358895657 | −4.121562668 |
| 4 | 1, -4 | $1f_{7/2}$ | −4.121562668 | −4.358895657 | 0, 5 | $0h_{9/2}$ | −4.517787398 | −4.358895657 |
| 1 | 2, -1 | $2s_{1/2}$ | −3.164540329 | −3.748920980 | 1, 2 | $1d_{3/2}$ | −4.116720149 | −3.748920980 |
| 2 | 2, -2 | $2p_{3/2}$ | −3.748920980 | −4.116720149 | 1, 3 | $1f_{5/2}$ | −4.354112751 | −4.116720149 |
| 3 | 2, -3 | $2d_{5/2}$ | −4.116720149 | −4.354112751 | 1, 4 | $1g_{7/2}$ | −4.513208345 | −4.354112751 |
| 4 | 2, -4 | $2f_{7/2}$ | −4.354112751 | −4.513208345 | 1, 5 | $1h_{9/2}$ | −4.623886008 | −4.513208345 |

**Table 4.** The energy eigenvalues (in $fm^{-1}$) of the Hellmann potential in units $\hbar = 2m = 1$.

| State | $\delta$ | $b = +1$ | | $b = -1$ | |
|---|---|---|---|---|---|
| | | NU | Ref. [13] | NU | Ref. [13] |
| 1s | 0.001 | -0.251500 | -0.250999 | -2.250500 | -2.24900 |
| | 0.005 | -0.257506 | -0.254963 | -2.252506 | -2.24501 |
| | 0.01 | -0.265025 | -0.259852 | -2.255025 | -2.24005 |
| 2s | 0.001 | -0.064001 | -0.063494 | -0.563001 | -0.561502 |
| | 0.005 | -0.070025 | -0.067353 | -0.565025 | -0.557550 |
| | 0.01 | -0.077600 | -0.071928 | -0.567600 | -0.552697 |
| 2p | 0.001 | -0.064000 | -0.063495 | -0.563000 | -0.561502 |
| | 0.005 | -0.070000 | -0.067377 | -0.565000 | -0.557541 |
| | 0.01 | -0.077500 | -0.072020 | -0.567500 | -0.552664 |
| 3s | 0.001 | -0.029280 | -0.028764 | -0.250502 | -0.249004 |
| | 0.005 | -0.035334 | -0.032457 | -0.252556 | -0.245111 |
| | 0.01 | -0.043003 | -0.036557 | -0.255225 | -0.240435 |
| 3p | 0.001 | -0.029279 | -0.028765 | -0.250501 | -0.249004 |
| | 0.005 | -0.035309 | -0.032480 | -0.252531 | -0.245103 |
| | 0.01 | -0.042903 | -0.036644 | -0.255125 | -0.240404 |
| 3d | 0.001 | -0.029388 | -0.028767 | -0.250833 | -0.249003 |
| | 0.005 | -0.035817 | -0.032526 | -0.254151 | -0.245086 |
| | 0.01 | -0.043825 | -0.036813 | -0.258269 | -0.240341 |
| 4s | 0.001 | -0.029280 | -0.016601 | -0.141129 | -0.139633 |
| | 0.005 | -0.035334 | -0.020077 | -0.143225 | -0.135819 |
| | 0.01 | -0.043003 | -0.023551 | -0.146025 | -0.131381 |
| 4p | 0.001 | -0.017128 | -0.016602 | -0.141128 | -0.139633 |
| | 0.005 | -0.023200 | -0.020098 | -0.143200 | -0.135811 |
| | 0.01 | -0.030925 | -0.023641 | -0.145925 | -0.131351 |
| 4d | 0.001 | -0.017189 | -0.016604 | -0.141314 | -0.139632 |
| | 0.005 | -0.023464 | -0.020142 | -0.144089 | -0.135796 |
| | 0.01 | -0.031356 | -0.023814 | -0.147606 | -0.131290 |
| 4f | 0.001 | -0.017311 | -0.016607 | -0.141686 | -0.139631 |
| | 0.005 | -0.024027 | -0.020206 | -0.145902 | -0.135772 |
| | 0.01 | -0.032356 | -0.024056 | -0.151106 | -0.131200 |

**Table 4.** continued

| State | δ | b = −2 | | b = −4 | |
|---|---|---|---|---|---|
| | | NU | Ref. [13] | NU | Ref. [13] |
| 1s | 0.001 | -4.000000 | -3.998000 | -8.999000 | -8.996000 |
| | 0.005 | -4.000006 | -3.990020 | -8.995006 | -8.980020 |
| | 0.01 | -4.000025 | -3.980070 | -8.990025 | -8.960100 |
| 2s | 0.001 | -1.000001 | -0.998000 | -2.249001 | -2.246000 |
| | 0.005 | -1.000025 | -0.990075 | -2.245025 | -2.230100 |
| | 0.01 | -1.000100 | -0.980297 | -2.240100 | -2.210400 |
| 2p | 0.001 | -1.000000 | -0.998002 | -2.249000 | -2.246000 |
| | 0.005 | -1.000000 | -0.990062 | -2.245000 | -2.230080 |
| | 0.01 | -1.000000 | -0.980248 | -2.240000 | -2.210330 |
| 3s | 0.001 | -0.444447 | -0.442451 | -0.999002 | -0.996009 |
| | 0.005 | -0.444501 | -0.434611 | -0.995056 | -0.980220 |
| | 0.01 | -0.444669 | -0.425103 | -0.990225 | -0.960885 |
| 3p | 0.001 | -0.444446 | -0.442451 | -0.999001 | -0.996008 |
| | 0.005 | -0.444476 | -0.434599 | -0.995031 | -0.980207 |
| | 0.01 | -0.444569 | -0.425055 | -0.990125 | -0.960820 |
| 3d | 0.001 | -0.444888 | -0.442450 | -0.999666 | -0.996007 |
| | 0.005 | -0.446651 | -0.434574 | -0.998317 | -0.980174 |
| | 0.01 | -0.448825 | -0.424959 | -0.996603 | -0.960691 |
| 4s | 0.001 | -0.250004 | -0.248012 | -0.561504 | -0.558516 |
| | 0.005 | -0.250100 | -0.240294 | -0.557600 | -0.542894 |
| | 0.01 | -0.250400 | -0.231150 | -0.552900 | -0.524055 |
| 4p | 0.001 | -0.250003 | -0.248011 | -0.561503 | -0.558515 |
| | 0.005 | -0.250075 | -0.240281 | -0.557575 | -0.542878 |
| | 0.01 | -0.250300 | -0.231103 | -0.552800 | -0.523991 |
| 4d | 0.001 | -0.250251 | -0.248010 | -0.561876 | -0.558514 |
| | 0.005 | -0.251277 | -0.240257 | -0.559402 | -0.542845 |
| | 0.01 | -0.252606 | -0.231011 | -0.556356 | -0.523865 |
| 4f | 0.001 | -0.250749 | -0.248009 | -0.562624 | -0.558512 |
| | 0.005 | -0.253714 | -0.240221 | -0.563089 | -0.542797 |
| | 0.01 | -0.257356 | -0.230872 | -0.563606 | -0.523674 |

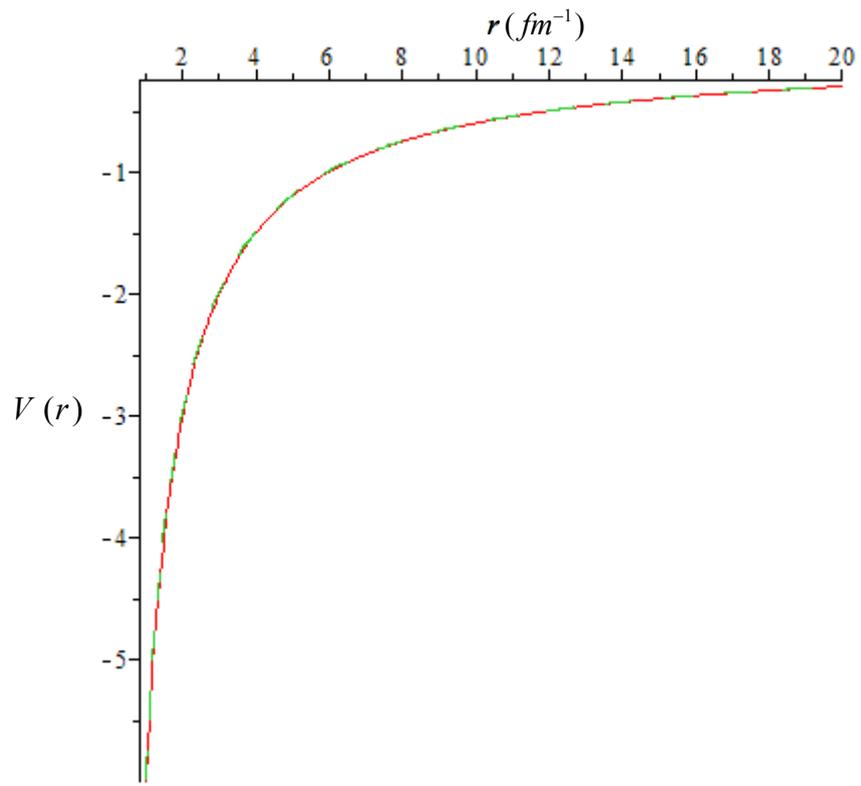

**Figure 1:** The Hellmann potential (red line) and its approximation in Eq. (20a) (green line).

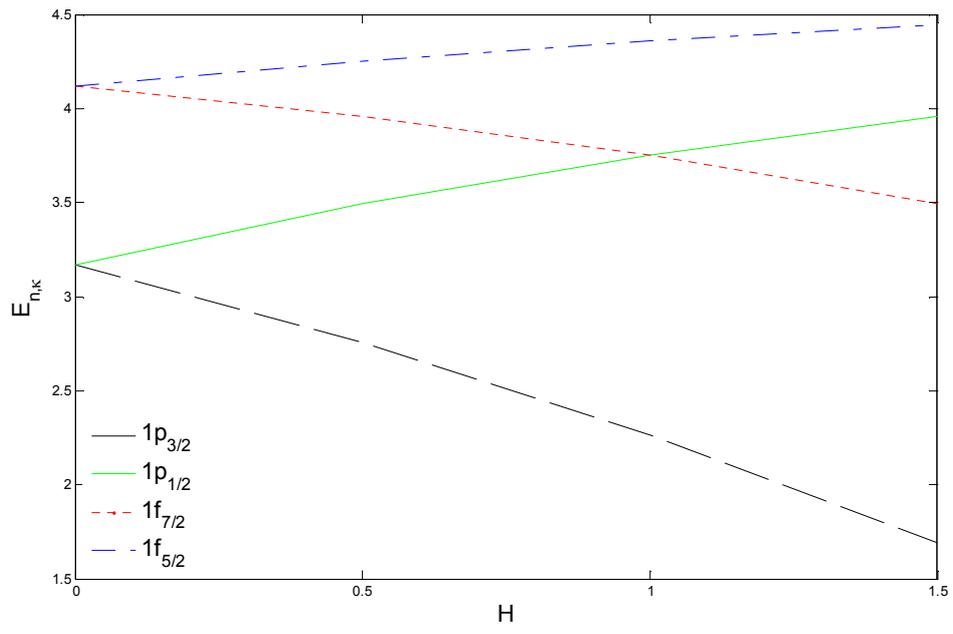

**Figure 2.** Effect of tensor potential on spin doublets

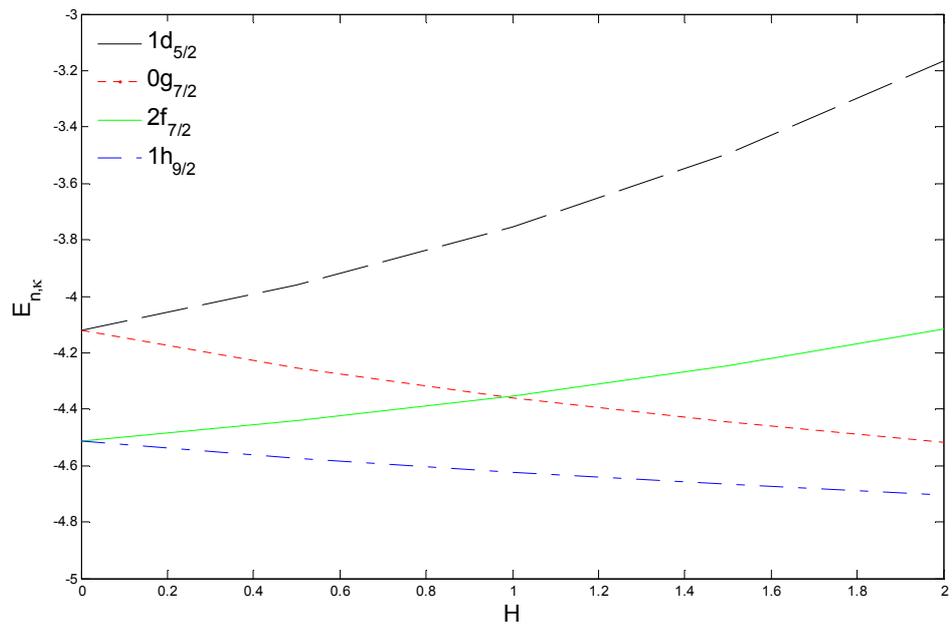

**Figure 3.** Effect of tensor potential on p-spin doublets

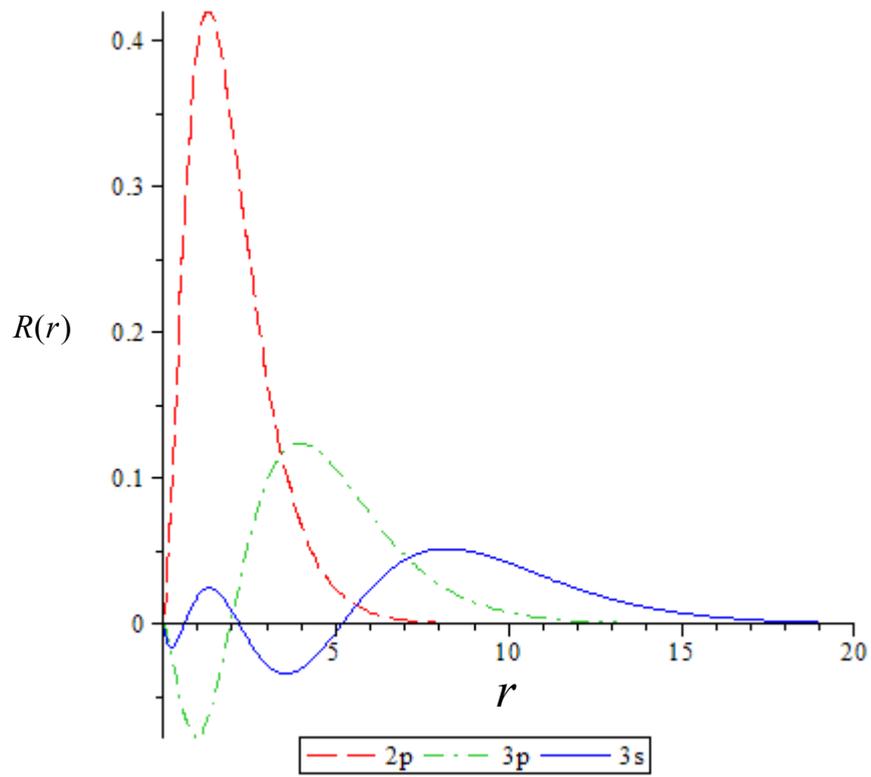

**Figure 4:** The normalized radial wave functions of the Hellmann potential for various states in the non-relativistic limit.